\documentclass[aps,prb,twocolumn,superscriptaddress]{revtex4}
\usepackage{amsmath,amsfonts,epsf}
\newcommand{\Tr}{\operatorname{Tr}}

\begin{document}


\title{Universal Quantum Signature of Mixed Dynamics in Antidot Lattices}

\author{J. P. Keating}
\affiliation{School of Mathematics, University of Bristol BS8 1TW,
UK}

\author{S. D. Prado}
\affiliation{Instituto de F\'{\i}sica, Universidade Federal do Rio
Grande do Sul, 91501-970 Porto Alegre, RS, Brazil}

\author{M. Sieber}
\affiliation{School of Mathematics, University of Bristol BS8 1TW,
UK}

\date{\today}

\begin{abstract}
We investigate phase coherent ballistic transport through antidot
lattices in the generic case where the classical phase space has both
regular and chaotic components. It is shown that the conductivity
fluctuations have a non-Gaussian distribution, and that their moments
have a power-law dependence on a semiclassical parameter, with
fractional exponents. These exponents are obtained from bifurcating
periodic orbits in the semiclassical approximation. They are
universal in situations where sufficiently long orbits contribute.

\end{abstract}


\maketitle

\section{Introduction}

Experiments on mesoscopic semiconductor devices have exhibited
a variety of features that can be attributed to quantum chaos
\cite{LesHouches:1995, ChaSolFract:1997, Nakamura:book}.
Modern fabrication techniques make it possible to produce extremely pure
semiconductor microstructures in which the motion of the electrons
is confined to two-dimensional domains whose relevant dimensions
are much smaller than the phase coherence length and the transport
mean-free-path. In this ballistic regime semiclassical
methods have been very successful in connecting quantum
interference effects to the underlying classical dynamics
\cite{Jalabert:2000, Richter:book}.

There has been a considerable emphasis on systems in which the confinement 
potential leads to chaotic motion. As in disordered systems,
certain transport properties are found to be universal, like 
conductance fluctuations and weak localization properties of
transport through cavities \cite{Bluemel:1988, Jalabert:1990,
Marcus:1992, Baranger:1993, Keller:1994, Chang:1994}. 
In contrast, if the dynamics is
integrable, these features are in general not universal but depend
on the specific system. Semiclassical methods have been applied to
explain universal properties of chaotic transport as well as the
non-generic behaviour of integrable cavities
\cite{Bluemel:1988, Jalabert:1990, Baranger:1993, Wirtz:1997,
Pichaureau:1999}.

Other experiments have been concerned with revealing signatures of classical
periodic orbits in quantum phenomena. Examples are orbital magnetism
in ballistic microstructures \cite{Richter:1996}
and transport through antidot superlattices \cite{Weiss:1991,
Fleischmann:1992, Weiss:1993, Hackenbroich:1995, Richter:1995}.
The latter consists of a two-dimensional electron gas at the 
interface of a $\text{GaAS/Al}_x\text{Ga}_{1-x}\text{As}$ heterostructure
into which a periodic array of holes is drilled. The effective potential
is a periodic structure of high potential peaks, and if the potential is
steep it may be considered as an experimental realization of the Sinai
billiard. Experiments on ballistic transport at temperature
$T \approx 1.5 K$ show a series of pronounced peaks in the longitudinal
conductivity vs magnetic field that can be explained by {\em classical}
electron transport within the Drude formalism
\cite{Weiss:1991, Fleischmann:1992}. Experiments
at a lower temperature $T \approx 0.4 K$ reveal additional
{\em quantum} oscillations superimposed on the classical peaks
\cite{Weiss:1993}. These quantum oscillations can be attributed
to unstable periodic orbits of the electrons in the confinement potential
by a semiclassical theory for the conductivity
\cite{Hackenbroich:1995, Richter:1995}.

Generic systems have a phase space which is neither completely chaotic
nor integrable but contains a mixture of regular islands and chaotic
regions. This is relevant for experiments because confinement potentials
are not hard-wall potentials,
and soft-wall potentials typically lead to mixed dynamics. A natural
question is whether ballistic mesoscopic systems have characteristic
properties that differ from those of chaotic and integrable systems.
One signature of mixed dynamics has been found in the ballistic transport
through cavities. It was predicted and observed experimentally that 
the variance of conductance fluctuations, as a function of the magnetic
field, has a power law dependence with a non-integer exponent that is
related to the trapping of chaotic trajectories near regular islands
\cite{Ketzmerick:1996, Sachrajda:1998}.
The purpose of this article is to show that there is a different
mechanism in ballistic transport through antidot lattices which leads
to {\em universal} behaviour in mixed systems.

Amongst the main characteristics of the dynamics in mixed systems are
bifurcations of periodic orbits, events in which different
periodic orbits coalesce when parameters of the system
are varied. Bifurcations are important for semiclassical approximations
because bifurcating orbits carry a semiclassical weight that is higher
than that of unstable periodic orbits and sometimes even that of
tori of regular orbits. The dominating influence of bifurcations
on transport through antidot lattices has been demonstrated in
\cite{Nihey:1995, Ma:2000}.
Bifurcations occur in different forms, and depending on the physical
quantity that one considers, they are of different
importance. In the following we consider moments of
conductivity fluctuations. We show that they are dominated by a
competition between different types of bifurcations, leading to
a power-law dependence on a semiclassical parameter
with fractional exponents. These exponents are universal if the
competition is amongst all generic bifurcations. Similar results have been
obtained for moments of spectral counting functions and wavefunctions
in closed mixed systems \cite{Berry:1998, Berry:2000, 
Berry_1:2000, Keating:2001}.

In the following we briefly review the semiclassical theory
of transport through antidot lattices and discuss modifications
in the presence of bifurcations. We give an overview of different
types of bifurcations and discuss their influence on moments
of conductivity fluctuations.

\section{Semiclassical theory for the conductivity}

The starting point for the semiclassical theory is the Kubo linear
response theory. The conductivity is given in terms of matrix elements
of the current operator. The semiclassical
approximation of the matrix elements\cite{Wilkinson:1987}
and the application of the stationary phase method
for evaluating integrals then yields a semiclassical expression
for the conductivity as a sum over the classical (Drude) component
\cite{Fleischmann:1992} and an oscillatory component in terms of
periodic orbits \cite{Hackenbroich:1995, Richter:1995}. 

The Kubo formula for the longitudinal conductivity is
\cite{Streda:1982}
\begin{equation}
\sigma_{xx} = \frac{e^2\pi\hbar}{V}{\mathrm{Tr}}
\left\{ \hat{v}_{x} \delta_{\Gamma}(E_F-\hat{H}) \hat{v}_{x}
\delta_{\Gamma}(E_{F}-\hat{H}) \right\} \label{Kubo}
\end{equation}
where $\hat{H} = (1/2m) (\hat{\mathbf{p}} - 
e \hat{\mathbf{A}}(\hat{\mathbf{r}}))^2 + U(\mathbf{r})$ is the Hamiltonian
of a two-dimensional electron gas in a perpendicular magnetic field,
${\mathbf{A}}=(-By/2,Bx/2)$ is the vector potential taken in the
symmetric gauge, $U(\mathbf{r})$ is the confinement potential, and 
$\hat{v}_x$ is the $x$ component of the velocity operator.  $E_F$ 
denotes the Fermi energy, $m$ is the effective mass of the electron,
and $V$ is the area of the system. Weak disorder is taken into 
account at the level of Born approximation by giving the $\delta$-functions
a finite width $\Gamma = \hbar / 2\tau_{el}$ where $\tau_{el}$
is the elastic scattering time.

To evaluate the Kubo formula (\ref{Kubo}) it is convenient to write
the delta function in terms of retarded and advanced Green functions:
$\delta_{\Gamma}(E_F - \hat{H}) = -(1/2\pi i) 
[\hat{G}^{+}(E_{F}) - \hat{G}^{-}(E_{F})]$
with $\hat{G}^{\pm}(E)=[E-\hat{H}\pm i\Gamma]^{-1}$. In the semiclassical
limit as $\hbar \rightarrow 0$, the longitudinal conductivity is
reduced to terms that involve only products of the retarded and advanced
Green functions, because terms that involve products of Green functions
of the same type vanish \cite{Hackenbroich:1995, Richter:1995}: 
\begin{align} \label{longcond}
\sigma_{xx} \approx & 2\frac{e^2\pi\hbar}{V}\left(
\frac{1}{2\pi }\right )^2 \int d^{2}r
d^{2}{r'}v_{x}({\mathbf{r}})v_{x}({\mathbf{r'}}) \times \\ \notag
& G^{-}_{{\mathbf{r}},{\mathbf{r'}}}(E_{F})G^{+}_
{{\mathbf{r'}},{\mathbf{{r}}}}(E_{F})  .
\end{align}
The Green function
$G^{+}_{({\mathbf{r'}},{\mathbf{{r}}})}(E)$ is semiclassically
approximated by a sum over classical trajectories $\gamma$ that
start at the initial position ${\mathbf{r}}$ and end up at the
final position ${\mathbf{r'}}$ with energy $E$ given
by \cite{Gutzwiller:1971, Gutzwiller:book}:
\begin{align} \label{semigreen}
G^{+}_{ \mathbf{r'},\mathbf{r} } & = \sqrt{ \frac{i}{2\pi\hbar^3} } 
\sum_{\gamma}  \exp \left( -\frac{T_\gamma}{2 \tau_{el} } \right)
\mid D_{\gamma} \mid^{1/2}
\times \notag \\ & \exp{\left [
i\frac{S_{\gamma}}{\hbar}({\mathbf{r'}},{\mathbf{r}})- i
\frac{\pi}{2}\eta_{\gamma} \right ]}.
\end{align}
$S_{\gamma}({\mathbf{r'}},{\mathbf{r}})=\int_{\gamma}{\mathbf{p}}\cdot
d{\mathbf{r}}$ is the classical action of the trajectory $\gamma$
with the canonical momentum ${\mathbf{p}}$, $\eta_{\gamma}$ is the
number of conjugate points along $\gamma$, $T_{\gamma}$ is the traversal
time of the trajectory and the first exponential is the weak disorder
factor. The amplitude factor $D_{\gamma}$ involves second derivatives
of $S_{\gamma}$. It is convenient to express it in terms of local
coordinates $\mathbf{r} = (z,y)$, with coordinate $z$ along the orbit
and $y$ perpendicular to it. Then  
\begin{equation}
D_{\gamma}= - \frac{1}{v v'} 
\frac{\partial^2 S_{\gamma}}{\partial y \partial y'}
\end{equation}
where $v$ and $v'$ are the velocities at $z$ and $z'$. Inserting
(\ref{semigreen}) into (\ref{longcond}) one finds that the longitudinal
conductivity is given by a double sum over oscillatory terms.
Due to the fact that $G^{+}(E)$ and $G^{-}(E)$ are complex conjugates,
$G^{-}_{{\mathbf{r}},{\mathbf{r'}}}(E)
= [G^{+}_{{\mathbf{r'}},{\mathbf{r}}}(E)]^{*}$,
the phase of the oscillatory terms contains the action difference
of two trajectories ($\gamma_1$ and $\gamma_2$). The conductivity 
is then split into two parts, a non-fluctuating one from the
diagonal terms ($\gamma_1=\gamma_2=\gamma$) and a fluctuating one
from the non-diagonal terms (with $\gamma_1 \neq \gamma_2$), that is:
\begin{equation}
\sigma_{xx}\approx
\overline{\sigma_{xx}}+\sigma_{xx}^{fl}. \label{cpm}
\end{equation}
The mean conductivity $\overline{\sigma_{xx}}$
\begin{equation}
\overline{\sigma_{xx}}=\frac{e^2}{ h^2 V} \int{d^2 r d^2
r'\sum_{\gamma ({\mathbf{r}},{\mathbf{r'}})}v_x({\mathbf{r}})
v_{x}({\mathbf{r'}})D_{\gamma}}\exp{(-T_{\gamma}/\tau_{el})},
\end{equation}
can be transformed into a familiar form as a phase-space average
(denoted by $\langle \cdot \rangle_{\mathbf{r,p}}$) where it
is recognized as the Drude conductivity \cite{Hackenbroich:1995,
Richter:1995}
\begin{equation}
\overline{\sigma_{xx}} = e^2\rho(E_F) \int^{\infty}_{0}dt
\langle v_x(0) v_{x}(t) \rangle_{\mathbf{r,p}}
\exp{(-t/\tau_{el})} \label{classical}
\end{equation}
where $\rho(E_F)=m/2\pi\hbar^2$ is the density of states per unit
area at the Fermi energy in two dimensional systems.

The fluctuating part $\sigma_{xx}^{fl}$ is a quantum correction
to the Drude conductivity and is given by the semiclassical formula
\begin{align} \label{fluct}
&\sigma_{xx}^{fl}=\frac{e^2}{h^2V}
\sum_{\gamma_1,\gamma_2} \int d^2r~d^2r'
(v_1)_x(v'_2)_x \sqrt{D_{\gamma_1}D_{\gamma_2}}
\notag \\
&\exp \left( -\frac{T_{\gamma_1,\gamma_2}}{2\tau_{el}} \right)
\exp \left( \frac{i}{\hbar}S_{\gamma_1,\gamma_2}({\mathbf{r}},
{\mathbf{r'}};E_F)-\frac{i\pi}{2} \eta_{\gamma_1,\gamma_2} \right),
\end{align}
where $T_{\gamma_{1},\gamma_2}=(T_{\gamma_{1}}+T_{\gamma_{1}})$;
$S_{\gamma_1,\gamma_2}= S_{\gamma_2}-S_{\gamma_1}$ and
$\eta_{\gamma_1,\gamma_2}=\eta_{\gamma_1}-\eta_{\gamma_2}$.
A correct evaluation of these integrals requires a detailed knowledge
of classical phase space structures as will become clear as we
proceed.

The main contribution to the integrals comes from the stationary
points where
\begin{align}
\nabla S_{\gamma_1,\gamma_2}({\mathbf{r'}},{\mathbf{r}};E_F)& =
{\mathbf{p}}_2-{\mathbf{p}}_1 = 0  \notag \\
\nabla'S_{\gamma_1,\gamma_2}({\mathbf{r'}},{\mathbf{r}};E_F)& =
{\mathbf{p'}}_1-{\mathbf{p'}}_1=0 
\end{align}
Here $\mathbf{p}_i$ and $\mathbf{p'}_{i}$ denote the initial and
final momenta of trajectory $\gamma_i$. The two trajectories
$\gamma_1$ and $\gamma_2$ must hence have the same initial and final
momenta in addition to having the same initial and final positions.
One way of satisfying these conditions with $\gamma_1$ and
$\gamma_2$ not being identical, is that both trajectories are part
of a primitive periodic orbit $\gamma$. In fact, for each periodic
orbit $\gamma$ there is an infinite set of pairs that satisfy
the stationary-phase conditions. They differ only by the number of
times they wind around $\gamma$ and their action difference is a
multiple of the action of the primitive periodic orbit
$S_{\gamma_2}-S_{\gamma_1}=r S_\gamma$.

The evaluation of (\ref{fluct}) is closely related to performing
the trace in the derivation of Gutzwiller's Trace Formula.
The integration is done in the local coordinates
${\mathbf{r}}=(z,u)$ with $d^2r d^2r'=dzdz'dy dy'$. In the case of
isolated periodic orbits the integrals over $y$ and $y'$ are
evaluated by
the stationary phase method while the integrals over $z$ and $z'$ are
performed exactly \cite{Hackenbroich:1995, Richter:1995}. This results in
\begin{align} \label{isolated}
\sigma_{xx}^{fl} = & \frac{2 e^2}{h V} \sum_\gamma \sum_{r=1}^\infty
\exp\left( \frac{-r T_\gamma}{2 \tau_{el}} \right)
C_\gamma(v_x,v_x) \notag \\
& \frac{\cos( r S_\gamma(E_F)/\hbar - \pi r \mu_\gamma/2)}{
| \det(M_\gamma - 1) |^{1/2}} \, ,
\end{align}
where $C_\gamma(v_x,v_x)$ is a correlation function of the longitudinal
components of the velocity along the orbit $\gamma$
\begin{equation}
C_\gamma(v_x,v_x) = \int_0^{T_\gamma} d t \int_0^\infty d t' \;
v_x(t) v_x(t+t') e^{-t'/\tau_{el}} \; .
\end{equation}
Formula (\ref{isolated}) applies to the fully chaotic case where
all orbits are unstable. In a lattice there are many copies of 
an orbit, so when counting different orbits one has to multiply
each one with a degeneracy factor, depending on the geometry.
Finite temperature and spin effects lead to additional factors of
$(\pi r T_\gamma/ \hbar \beta) / \sinh(\pi r T_\gamma / \hbar \beta)$
and $2 \cos(r T_\gamma \mu_B B / \hbar)$, respectively, where
$\mu_B = e \hbar / 2 m_e$ is the Bohr magneton.
There are similar formulas for the Hall conductivity $\sigma_{xy}$
\cite{Hackenbroich:1995, Richter:1995, Nakamura:book}.

Close to a bifurcation the saddle point approximation leading to
(\ref{isolated}) breaks down because the saddle point is not isolated.
There are nearby saddle points from the other periodic orbits that
participate in the bifurcation.
This occurs when, by changing the energy or parameters of the system,
the eigenvalues of $M^r \rightarrow 1$ and (\ref{isolated}) diverges.
In order to obtain the correct semiclassical contribution one has to
integrate over the neighbouring saddle points as well. This results in 
transitional or uniform approximations for the bifurcating orbits.

Bifurcations occur in specific forms that depend on the repetition
number $r$ for which $\det M^r = 1$. They are characterized by
normal forms which describe the characteristic motion of trajectories
in the vicinity of a periodic orbit
\cite{Ozorio:book}. Before describing the normal
forms we transform the integral in (\ref{fluct}) into a form that
is more appropriate for treating bifurcations.

One of the integrals over $y$ and $y'$ is done in stationary phase
approximation, and the other is transformed to an integral over
normal form coordinates \cite{Ozorio:1987,Sieber:1996}.
Afterwards the integral over the $z$ and
$z'$ coordinates can be performed. The resulting contribution
from the $r$-th repetition of a bifurcating orbit $\gamma$ is

\begin{equation} \label{transi}
\frac{e^2 \exp\left( - \frac{r T_\gamma}{2 \tau_{el}} \right)
}{2 \pi^2 V} C_\gamma(v_x,v_x) \Re \left[
\exp( \frac{i}{\hbar} r S_0 - i \frac{\pi}{2} \nu_r)
\mathcal{G}(E_f) \right]
\end{equation}
where
\begin{equation} \label{ge}
\mathcal{G}(E_f) = \frac{1}{\hbar^2} \int d Q' \, d P \; 
\exp\left( \frac{i}{\hbar} \Phi(Q',P) \right) \; .
\end{equation}

Here $S_0$ is the action of the periodic orbit at the centre of the
Poincar\'e surface, and $\nu_r$ is the number of conjugate points
along $r$ repetitions of this orbit. $\Phi(Q'P)$ is a generating
function for the Poincar\'e map in normal form coordinates from
$(Q,P)$ to $(Q',P')$. At the position of a periodic point
\begin{equation}
\frac{\partial \Phi}{\partial Q'} = P' - P = 0 \; , \qquad
\frac{\partial \Phi}{\partial P } = Q - Q' = 0 \; ,
\end{equation}
and $\Phi$ is stationary.

Eq. (\ref{transi}) is a transitional approximation for the bifurcating
orbits. It is correct when the orbits are close to a bifurcation, where
its semiclassical effect is largest \cite{Ozorio:1987}. 
If, due to a change of parameters,
the orbits move further apart, it reduces to a sum of single terms
as in (\ref{isolated}) where, however, the amplitudes for the neighbouring
orbits are inaccurate. The transitional approximation is sufficient
for our purpose. A uniform approximation which has the correct single
orbit limit, can be obtained by including the correlation
function $C_\gamma(v_x,v_x)$ into the integral (where it is evaluated
along neighbouring trajectories) and including also the $Q'$ and $P$
dependence of the pre-exponential factor that comes from the
determinants $D$ \cite{Sieber:1996,Schomerus:1997}. In (\ref{transi})
this factor has been evaluated at the central orbit and has the value one.

Let us look at an example of a pitchfork bifurcation with normal form
\begin{equation} \label{pitch}
\Phi(Q,P) = \frac{P^2}{2} + \varepsilon Q^2 + c Q^4 \; ,
\end{equation}
where $c>0$ and the primes have been dropped for convenience.
For $\varepsilon > 0$ there is only one solution of 
$0 = \frac{\partial \Phi}{\partial Q} = \frac{\partial \Phi}{\partial P}$
at $(0,0)$ corresponding to the central periodic orbit. If, by
changing parameters, $\varepsilon$ goes through zero, two new
solutions appear which are located symmetrically about $Q=0$.
In the generic situation (in systems without symmetries) this 
is a period doubling bifurcation. It occurs if the lowest
repetition number $r$ for which $\det(M^r - 1) = 0$ is $r=2$.
At the bifurcation one new orbit of double the period arises
which intersects the Poincar\'e section twice.

The integral (\ref{ge}) can be evaluated in closed form for
the normal form (\ref{pitch}) and is given by a sum of two
Bessel functions with index $\frac{1}{4}$ and $-\frac{1}{4}$.
At the bifurcation, $\varepsilon=0$, it has the simple form
\begin{equation} \label{pitchres}
\mathcal{G}(E) = 
\sqrt{\frac{\pi}{8}} \frac{\Gamma(1/4) \exp(3 \pi i / 4)
}{\hbar^{5/4} \, c^{1/4}}
\end{equation}
The exponent of $\hbar$ in the denominator, $\beta = 5/4$, 
shows that the bifurcation term is by a factor $\hbar^{-1/4}$
stronger than that of a single orbit. The contribution
of pitchfork and tangent bifurcations to the conductivity
were evaluated in \cite{Ma:2000,Nakamura:book}.

We note that the constant $c$ in the normal form that appears
in (\ref{pitchres}) can be obtained by following the actions
and stabilities of the orbits through the bifurcations.
It follows from \cite{Schomerus:1997}
\begin{equation}
\Tr M_0 - \Tr M_1 \sim - 6 \varepsilon \, , \qquad
S_0 - S_1 \sim \frac{\varepsilon^2}{4 c} \; .
\end{equation}
Hence one finds
\begin{equation}
c = \lim_{\varepsilon \rightarrow 0} \frac{(\Tr M_0 - \Tr M_1)^2}{
144 (S_0 - S_1)} \; .
\end{equation}

There is a second exponent, $\gamma$, which characterizes the 
semiclassical importance of bifurcations for spectral and
transport properties. It specifies the region in parameter
space ($\varepsilon$ in the example) in which the contribution
of the bifurcation is strongest. Both $\beta$ and $\gamma$ can
be obtained from the normal form by a scaling argument. We
perform a scaling of $Q$, $P$ and $\varepsilon$ such that the
integral in (\ref{ge}) takes the form

\begin{equation} \label{scaling}
\mathcal{G}(E_f, \varepsilon, \hbar)=
\frac{1}{\hbar^\beta} \mathcal{G} \left (
E_f, \frac{\varepsilon}{\hbar^{\gamma}}, 1 \right ).
\end{equation}

In the example of the pitchfork bifurcation the argument of 
the exponent is made $\hbar$-independent by changing the
integration variables and a subsequent scaling of $\varepsilon$.
\begin{equation}
P = \hbar^{1/2} \tilde{P} \, , \quad
Q = \hbar^{1/4} \tilde{Q} \, , \quad
\varepsilon = \hbar^{1/2} \tilde{\varepsilon} \, .
\end{equation}

As a result $\beta=2- 1/2 - 1/4 = 5/4$, as before, and $\gamma=1/2$.

Different bifurcations have different exponents $\beta$ and $\gamma$
and, depending on the quantities that one considers, they
are of different importance. We will
consider in the following statistical properties of the conductivity.
It will be shown that they are dominated by certain kinds of
bifurcations, leading to universal properties in the regime of 
mixed classical dynamics. Before we do that we give an overview 
of classifications of periodic orbits that are known presently.

The generic bifurcations of codimension $K=1$ are those that occur
if one parameter of the system is varied. They have been classified
by Meyer and Bruno \cite{Meyer:1970,Brjuno:1970}.
A list of the normal forms can be obtained
from Table \ref{table_1} by setting $x_2=1$ (except for $r=1$ 
where it is $P^2 + x_1 Q + Q^3$).
These normal forms are simplified versions in which all constants
and terms that are irrelevant for the determination of $\beta$ and
$\gamma$ have been removed. The scaling proceedure of Eq. (\ref{scaling})
yields the exponents $\beta_r$ and $\gamma_r$ that are given in 
Table \ref{table_2}.

If more parameters than one are varied then other bifurcations occur
in which more complicated configurations of periodic orbits coalesce.
For example, by varying a second parameter one can make certain
codimension $K=1$ bifurcations occur at the same instance.
The codimension $K$
corresponds to the number of parameters that are varied. Generic
bifurcations of codimension $K=2$ have been classified by Schomerus,
\cite{Schomerus:1998}
and the normal forms are given in Table \ref{table_1}. 
For each parameter $x_j$ one
has an exponent $\sigma_j$ that specifies the range over which the
bifurcation is important, $\Delta x_j \sim \hbar^{\sigma_j}$. The total
volume in $K$-dimensional parameter space then scales as $\hbar^{\gamma}$ where
\begin{equation}
\gamma = \sum_{j=1}^K \sigma_j \; .
\end{equation}

The exponents $\beta_{K,r}$ and $\gamma_{K,r}$ for bifurcations of
codimension $K=2$ are also listed in Table \ref{table_2}. 

\begin{table}
\vspace{0.3cm}
\begin{ruledtabular}
\begin{tabular}{cc}
  r& $\Phi_{r,2}$ \\ \hline
  1 & $P^2 + x_1 Q + x_2 Q^2 + Q^4$ \\ 
  2 & $P^2 + x_1 Q^2 + x_2 Q^4 + Q^6$ \\
  3 & $(P^2 + Q^2)^2 + x_1 (P^2+Q^2) + x_2 {\mathrm Re}[(P+iQ)^3]$ \\
  4 & $P^2 Q^2 + x_1 (P^2+Q^2) + x_2 (P^2-Q^2)^2 $ \\
  5 & $\mathrm{Re}[(P + iQ)^5] + x_1(P^2+Q^2) + x_2(P^2+Q^2)^2$ \\
  $\geq 6$ & $(P^2+Q^2)^3 + x_1(P^2+Q^2) + x_2(P^2+Q^2)^2$
\end{tabular}
  \caption{The relevant parts of the normal forms for
bifurcations of period-$r$ orbits with codimension $K=2$.}
  \label{table_1}
  \end{ruledtabular}
\end{table}

\begin{table}
  \begin{ruledtabular}
\begin{tabular}{ccccc}
  r        & $\beta_{1,r}$ & $\gamma_{1,r}$ & $\beta_{2,r}$ & $\gamma_{2,r}$\\
\hline
  1        & $\frac{7}{6}$ & $\frac{2}{3}$ & $\frac{5}{4}$ & $\frac{5}{4}$\\
  2        & $\frac{5}{4}$ & $\frac{1}{2}$ & $\frac{4}{3}$ & ${1}$\\
  3        & $\frac{4}{3}$ & $\frac{1}{3}$ & $\frac{3}{2}$ & $\frac{3}{4}$\\
  4        & $\frac{3}{2}$ & $\frac{1}{2}$ & $\frac{3}{2}$ & $\frac{1}{2}$\\
  5        & $\frac{3}{2}$ & $\frac{1}{2}$ & $\frac{8}{5}$ & $\frac{4}{5}$\\
  $\geq$6  & $\frac{3}{2}$ & $\frac{1}{2}$ & $\frac{5}{3}$ & $1$
\end{tabular}
  \caption{Exponents $\beta_{K,r}$ and $\gamma_{K,r}$ for generic
bifurcations with codimension $K=1$ or $K=2$.}
\label{table_2}
  \end{ruledtabular}
\end{table}

Although bifurcations of higher codimension do not occur, in general,
when only one parameter is varied they still affect semiclassical
approximations, because of their finite extention in parameter space.
This is one of the main reasons why the semiclassical analysis of mixed
systems is so intricate.

For bifurcations of codimension $K \geq 3$ there is no complete
classification. Only partial results exist. 
Our main interest lies in bifurcations of arbitrary codimension $K$
with repetition number $r \geq 2 K + 2$. These have the (simplified)
normal form 
\begin{equation} 
\Phi_{r,K}(Q,P)=I^{K+1}+\sum_{n=1}^{K}x_nI^{n}+\mathcal{O}(I^{K+2}),
\end{equation}
where $I=Q^2+P^2$. Expressing $\Phi$ in terms of $Q$ and $P$, we find
\begin{align} \label{largeK}
\beta_{K,r} & =2 - \frac{2}{2(K+1)} = \frac{2K+1}{K+1} \, , \notag \\
\sigma_{K,r,n} & = 1 - \frac{n}{K+1} \quad \Longrightarrow \quad
\gamma_{K,r} = \sum_{n=1}^K \sigma_{K,r,n} = \frac{K}{2} \, .
\end{align}
It will be shown that these bifurcations are the most important for
determining conductance fluctuations in the next section.

\section{Moments of the conductivity fluctuations}

We estimate the semiclassical size of the conductivity fluctuations
$\sigma_{xx}^{fl}$ by evaluating the $\hbar$-dependence of the moments
\begin{equation}
M_{2m}(\hbar) = \left\langle [\sigma_{xx}^{fl}(E,\hbar)]^{2m} \right\rangle
\end{equation}
where $\langle \cdot \rangle$ denotes a local average over one or
more parameters of the Hamiltonian, for example Fermi energy, magnetic
field or system parameters. In this section we assume that very
long orbits do contribute to the longitudinal conductivity.
The consequence of the damping of the contributions of long
orbits due disorder and inelastic processes is discussed in
the next section. The central point now is to replace the
average by an average over the parameters in the normal forms
\begin{equation}
M_{2m,r,K}(\hbar) \equiv B  \int d^{K}{\bf{x}}
[\mathcal{C}^{fl}_{r,K}({\bf{x}},\hbar)]^{2m} \, ,
\end{equation}
where $B$ is a normalization constant.
The scaling proceedure in (\ref{scaling}) yields that each
bifurcation, labelled by $r$ and $K$ contributes a term that scales
as $1 / \hbar^{2 m \beta_{K,r} - \gamma_{K,r}}$. With the $\hbar$
dependence thus extracted, these contributions can now be compared
for different bifurcations. The bifurcation that wins the competition 
is that for which the $\hbar$ exponent in the denominator is largest,
and it determines that rate at which the $\mathbf{x}$-averaged
moments diverge in the semiclassical limit. That is
\begin{equation}
M_{2m}(\hbar) \sim \hbar^{-\nu_m} \quad \mathrm{as} 
\quad \hbar \rightarrow 0,
\end{equation}
where
\begin{equation}
\nu_m = \max_{K,r} ( 2 m \beta_{K,r} - \gamma_{K,r}) \, .
\end{equation}

The exponents $\nu_m$ are universal numbers that are determined by
studying the hierarchy of bifurcations. Similar universal exponents
have also been found for moments of the fluctuating parts of the
spectral counting function and wavefunctions
\cite{Berry:2000,Berry_1:2000,Keating:2001}. They have been named
`twinkling exponents', in analogy to the exponents that control
the intensity of twinkling starlight.

For the generic bifurcations with $K=1$ and $K=2$ the exponents
$\nu_{m,K,r} = 2 m \beta_{K,r} - \gamma_{K,r}$ are listed in
table \ref{table_3}. For generic bifurcations with $K > 2$ and
$r \geq 2 K + 2$ the exponents $\nu_{m,K,r}$ follow from 
Eq. (\ref{largeK}) and are given by
\begin{equation} \label{individ_nu}
\nu_{m,K,r} = 2 m \frac{2 K + 1}{K + 1} - \frac{K}{2} \, ,
\end{equation}
which are independent of $r$.
Although the normal forms for $K>3$ and $1 \leq r < K$ have
not been classified completely, it has been argued in refs.
\cite{Berry_1:2000, Keating:2001} that these bifurcations
cannot contribute to the twinkling exponents since they have a
counterpart with $\tilde{K} < K$ and 
$\tilde{r} > 2\tilde{K} + 2$ with the property
that $\nu_{m,\tilde{K},\tilde{r}} > \nu_{m,K,r}$. The same
argument can be applied in the present case, and hence the maxima of
the exponents given in (\ref{individ_nu}) represent the maxima
with respect to all generic bifurcations and are universal.
They can be written as
\begin{equation}
\nu_m = 4 m - \min_K \left( \frac{2 m}{K + 1} + \frac{K}{2} \right) \, .
\end{equation}
The exponents $\nu_m$ are given in Table \ref{table_4}.
If $m$ is a square then $\nu_m = 4 m - 2 \sqrt{m}
+ 1/2$. If not, the maximum value is attained at one or both of the
two integer values of $K$ that are closest to $2 \sqrt{m} - 1$. 
It is clear from the moments that the conductivity fluctuations do not
have a Gaussian distribution. 

\begin{table}
\begin{ruledtabular}
\begin{tabular}{ccccc}
  r        & $\nu_{1,1,r}$ & $\nu_{2,1,r} $ & $\nu_{1,2,r} $ & $\nu_{2,2,r} $\\
\hline
  1        & $\frac{5}{3}$ & $\frac{12}{3}$ & $\frac{5}{4} $ & $\frac{15}{4}$\\
  2        & $2          $ & $\frac{9}{2} $ & $\frac{5}{3} $ & $\frac{13}{3}$\\
  3        & $\frac{7}{3}$ & $5           $ & $\frac{9}{4} $ & $\frac{21}{4}$\\
  4        & $\frac{5}{2}$ & $\frac{11}{2}$ & $\frac{5}{2} $ & $\frac{11}{2}$\\
  5        & $\frac{5}{2}$ & $\frac{11}{2}$ & $\frac{12}{5}$ & $\frac{28}{5}$\\
  $\geq$6  & $\frac{5}{2}$ & $\frac{11}{2}$ & $\frac{7}{3} $ & $\frac{17}{3}$
\end{tabular}
  \caption{Values of $\nu_{m,K,r}=2m \beta_{K,r} - \gamma_{K,r}$.}
\label{table_3}
  \end{ruledtabular}
\end{table}

\begin{table}
\begin{ruledtabular}
\begin{tabular}{ccccccccc}
  m & 1 & 2 & 3 & 4 & 5 & 6 & 7 & 8 \\ \hline
  $\nu_m$ & $2 \frac{1}{2}$ & $5 \frac{2}{3}$ & 9 & $12\frac{1}{2}$ & 16 
  & $19\frac{3}{5}$ & $23\frac{1}{5}$ & $26\frac{5}{6}$ \\
   $K$ & 1 & 2 & 2,3 & 3 & 3,4 & 4 & 4 & 5 \\
\end{tabular}
\caption{Exponents $\nu_m$ and codimensions $K$ of
the dominanting bifurcations for generic two-dimensional systems.}
\label{table_4}
\end{ruledtabular}
\end{table}

Another quantity explored in the literature concerns the magnetic
fingerprint in antidot lattices incorporated in the auto correlation
function of the conductivity \cite{Nakanish:1996}. Consider
\begin{equation} \label{correlation}
\mathcal{F}(\Delta x) = \left\langle
\sigma_{xx}^{fl}(x+\Delta x) \times \sigma_{xx}^{fl} (x)
\right\rangle_{x,y}
\end{equation}
where $\sigma^{fl}_{xx}$ is written as a sum over classical orbits
and $x$ is a parameter of the system, e.g. the magnetic field.
The average $\langle \cdot \rangle_{x,y}$ might be over other
parameters as well. At $\Delta x = 0$ this is identical to the
second moment $M_2$. The bifurcations then determine
the characteristic length scale over which the correlations decay
as $\Delta x$ is increased. In the generic situation all 
parameters $x_n$ in the normal form are affected when the
physical parameter $x$ is changed. Hence the correlation
length scales as $\hbar^\sigma$ where $\sigma$ is the
minimum of the $\sigma_i$. For generic bifurcations with
$r \ge 2 K + 2$ this yields $\sigma = 1/(K+1)$.

\section{Discussion}

The semiclassical analysis of the conductivity fluctuations
leads to a powerlaw dependence of the moments $M_{2 m} \sim
1/\hbar^{\nu_m}$, with fractional exponents $\nu_m$.
In situations where arbitrarily long periodic orbits
contribute to the conductivity these exponents are
{\em universal} numbers that are obtained from the
competition between different bifurcation. The analysis
was done in terms of $\hbar$. If a different semiclassical
variable is used instead of $\hbar$, e.g. the Fermi wave
length, then the exponents have to be adjusted depending
on how the classical action scales with the semiclassical
parameter.

In experiments most of the long orbits are suppressed and
the conductivity is determined by a relatively small number
of periodic orbits. One cannot expect then, in general, 
to see the
universal exponents in particular for the higher moments,
because they originate from high repetitions of periodic
orbits. Instead the exponents will be determined by the
competition within the much smaller class of those bifurcations
that affect the relevant periodic orbits. The dominating
exponent can then be found by comparing the corresponding
exponents $\nu_{m,K,r}$ in Table \ref{table_3} (assuming
that higher codimensions are not important). Nevertheless,
also in this case the conductivity fluctuations would be
non Gaussian and the moments would have a powerlaw dependence
with fractional exponents in clear contrast to the chaotic
case with Gaussian fluctuations and $\nu_m = 2 m$. Also
the length scale of the decay of correlations is larger
if the classical dynamics is mixed, and is of order
$\hbar^\sigma$ with $\sigma<1$.

There is, however, one property of antidot lattices that might
make it possible to observe the {\em universal} exponents for
the lower moments, and that is their high symmetry. In systems
with discrete symmetries there are also non-generic bifurcations
which typically have the same normal form as generic bifurcations,
but at lower repetition numbers. For example, the pitchfork
bifurcation Eq. (\ref{pitch}) can occur at repetition number
$r=1$ where two new orbits of the same length appear instead
of one orbit of double the length \cite{Aguiar:1987}. 
Similarly, if the antidot lattice is e.g. invariant under
rotations of $\pi/3$ then a bifurcation that is generic for
$r=6$ can occur at the first repetition of a periodic orbit
\cite{Then:1999}. This might make it feasible to observe the
universal regime that in non-symmetric systems is restricted
to very long orbits.
\bigskip

\section*{Acknowledgements}

We are grateful to the Royal Society for funding this work.
JPK is supported by an EPSRC Senior Research Fellowship.
SDP thanks FAPERGS for partial support.

\end{document}